\documentclass[letterpaper,twocolumn,10pt]{article}
\usepackage{usenix-2020-09}

\usepackage{amsmath,amssymb}
\usepackage{booktabs}
\usepackage{array}
\usepackage{longtable}
\usepackage{tabularx}
\usepackage{graphicx}
\usepackage{float}
\usepackage{listings}
\usepackage{caption}
\usepackage{enumitem}
\usepackage{tikz}
\usetikzlibrary{arrows.meta,fit,positioning}
\usepackage{url}
\usepackage{fancybox}

\lstdefinestyle{routercode}{
  language=C,
  basicstyle=\ttfamily\small,
  frame=single,
  numbers=left,
  numberstyle=\tiny,
  breaklines=true,
  tabsize=4,
  showstringspaces=false,
  columns=fullflexible,
  keepspaces=true,
  xleftmargin=1em,
  framexleftmargin=1.5em,
  captionpos=b
}

\date{}
\title{\Large \bf IsoRoute: Isolating Route Installation\\for Scalable Container-Based Network Emulation}

\author{
{\rm Yi Liu$^{1}$} \quad {\rm Tian Pan$^{1}$} \quad {\rm Qiang Fu$^{2}$} \quad {\rm Guohao Ruan$^{1}$} \quad {\rm Tao Huang$^{1}$}\\
{\rm $^{1}$Beijing University of Posts and Telecommunications}\\
{\rm $^{2}$RMIT University}\\
{\rm \{liu\_yi, pan, ghruan, htao\}@bupt.edu.cn} \quad {\rm qiang.fu@rmit.edu.au}
}

\begin{document}

\maketitle

\begin{abstract}
Container-based emulation lets researchers run real routing software for thousands of nodes on a single machine, and it is a common way to study low Earth orbit (LEO) satellite networks. At constellation scale, besides the topology-construction problem that prior work has studied, a bottleneck appears on the runtime path that installs already computed routes. On our testbed, the route-update load generated by LEO dynamics grows quickly with the number of emulated satellites, while the throughput of kernel route installation instead falls by a factor of about 23 between 100 and 1,600 containers. We trace this decline to coupling between containers that namespaces do not remove: beyond the known lock problem in the kernel networking subsystem, the kernel scans, inside the lock, a Netlink subscriber list shared by all namespaces.

We present IsoRoute, which makes each emulated node an independent contention domain for forwarding-state updates. FRR exports routes through its Forwarding Plane Manager (FPM) interface to an agent on the host, and the agent writes them into each node's own eBPF longest-prefix-match table; a tc program forwards from these tables and bypasses only the FIB lookup of kernel forwarding. Traffic from applications inside a container reaches the same program through a default route, while everything else still goes through the kernel, so tools such as \texttt{ping} and \texttt{traceroute} keep working without modification. IsoRoute sustains about $3\times10^5$ route installations per second from 500 to 3,000 FRR containers, and reduces the OSPF data-plane convergence time of a 1,584-node Starlink topology from 594.7\,s to 49.5\,s. Our implementation is publicly available.

\end{abstract}

\section{Introduction}

Researchers need platforms that capture real routing behavior, protocol interactions and forwarding effects while the topology keeps changing. Container-based network emulation fits this need: it runs on commodity servers, costs far less per node than virtual machines, and, unlike simulation, executes the same routing software that runs on real routers.

LEO networks put unusual pressure on the runtime of such an emulator. Satellite motion keeps switching inter-satellite links on and off, and every change forces the control plane to compute and install new routes. Existing emulation work has mostly optimized how quickly a topology can be built~\cite{starrynet,opensn,splitnn}. We find that, at constellation scale, a hard limit appears at a later stage, when the emulation runs: computed routes cannot be installed as fast as the topology generates them.

\begin{figure}[t]
\centering
\includegraphics[width=\columnwidth]{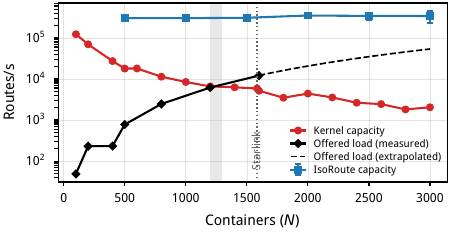}
\caption{Route-installation capacity and LEO offered load versus emulation scale. Load beyond 1,600 nodes is extrapolated with a power-law fit ($\propto N^{2.35}$, dashed); the shaded band marks the crossing.}
\label{fig:scale-sweep}
\end{figure}

Figure~\ref{fig:scale-sweep} shows this limit on our testbed. The route-update load grows superlinearly with constellation size, while the throughput of the standard kernel installation path falls; the two curves cross at about 1,200--1,300 containers. Beyond the crossing, updates arrive faster than they can be installed, the backlog keeps growing, and the data plane falls behind the control plane. Starlink Shell-1, with 1,584 satellites, is already past the crossing, and Kuiper Phase~1, with 3,236 satellites, is far beyond it; the numbers are in \S\ref{sec:barrier}.

Profiling and isolation experiments (\S\ref{sec:rootcause}) show where the time goes. Every route update serializes on the global \texttt{rtnl\_mutex}, a known problem~\cite{pernetns-rtnl}. We further observe, however, that inside this critical section the kernel scans a Netlink multicast subscriber list that is not partitioned by namespace. Every routing daemon adds sockets to this list, so the cost of one update in one container grows with the number of daemons on the host, and even idle daemons have an effect. The mutex prevents updates from running in parallel and caps performance; on its own, it would produce an almost flat performance curve that depends only weakly on the emulation scale. The scanning mechanism we identify is what makes each update slower as the emulation grows.

These observations lead to the design principle behind IsoRoute: forwarding-state updates of different emulated nodes should be isolated from each other, not implicitly coupled. Isolation, however, is not the only requirement. The design must also be usable on existing emulation platforms without changing the kernel and with almost no change to FRR, and the emulated network itself must stay the same: every node still forwards hop by hop, and \texttt{ping} and \texttt{traceroute} keep working. Satisfying any two of these is straightforward, but the obvious alternatives---patching the kernel, forwarding with DPDK, or sharing one eBPF table---each violate one of them (\S\ref{sec:design-requirements}).

IsoRoute keeps the control plane unchanged and replaces only the path that turns route decisions into forwarding state. \texttt{zebra} exports routes through FRR's Forwarding Plane Manager (FPM) interface to an agent on the host; the agent writes them into one eBPF longest-prefix-match table per node; and a tc program forwards packets from these tables. IsoRoute bypasses only the FIB lookup of kernel forwarding: transit packets consult the eBPF table at link ingress, and traffic from applications inside a container enters the same program through a default route to \texttt{dummy0}, while local delivery, ICMP generation and neighbor resolution remain with the kernel. Standard tools such as \texttt{ping} and \texttt{traceroute} therefore keep working without modification.

This paper makes the following contributions:
\begin{itemize}[leftmargin=1.2em,itemsep=1pt,topsep=2pt]
\item We identify the cause of the runtime scaling limit of container-based LEO emulation: a per-update scan of a host-wide Netlink subscriber list inside \texttt{rtnl\_mutex}, which couples containers that should be independent (\S\ref{sec:rootcause}).
\item We state three design requirements---node isolation, deployability, and emulation invariance---show why the obvious alternatives each violate one of them, and design, implement and open-source IsoRoute, which meets all three. It keeps an independent eBPF forwarding table per node, written by a sharded agent so that installations at different nodes do not interfere; it changes no kernel code and only one line of \texttt{zebra}; and it bypasses only the kernel's FIB lookup, leaving the rest to the kernel so that forwarding stays identical to hop-by-hop kernel forwarding (\S\ref{sec:design}).
\item Our evaluation shows that IsoRoute sustains about $3\times10^5$ route installations per second from 500 to 3,000 containers, scales to 16 cores, and reduces the OSPF data-plane convergence time of a 1,584-node Starlink topology from 594.7\,s to 49.5\,s (\S\ref{sec:evaluation}).
\end{itemize}

\section{Background and Problem}
\label{sec:background}

\subsection{Container-Based Network Emulation}

Container-based network emulation is widely used to evaluate routing protocols at scale~\cite{mininet,starrynet,opensn}. Unlike virtual machines, containers share the host kernel while providing isolated network namespaces, enabling researchers to emulate hundreds or thousands of routing nodes on a single physical machine with acceptable resource overhead.

\begin{figure}[t]
\centering
\includegraphics[width=\columnwidth]{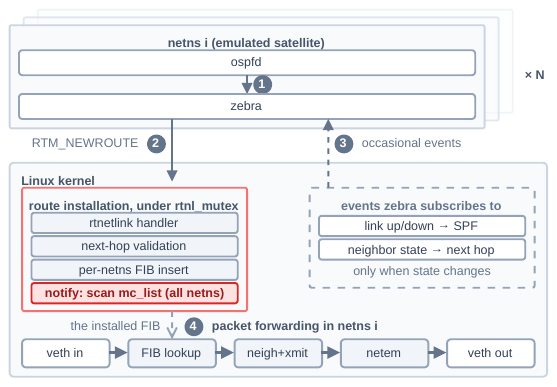}
\caption{Route installation and packet forwarding in the baseline. Red: the subscriber scan inside \texttt{rtnl\_mutex}, shared by all $N$ namespaces (\S\ref{sec:rootcause}).}
\label{fig:baseline-path}
\end{figure}

Each emulated node runs a routing stack (e.g., \texttt{ospfd} for OSPF) inside its own container. Routing daemons compute routes and push them to \texttt{Zebra}, which installs them into the kernel FIB via Netlink. Each container has an isolated FIB, but the installation path---\texttt{rtnl\_mutex} and Netlink notification---is shared by all containers. Figure~\ref{fig:baseline-path} shows this baseline: routes go from \texttt{ospfd} to \texttt{zebra} and over Netlink into the kernel FIB of the calling namespace; the kernel delivers link and neighbor events back to \texttt{zebra} (link events drive OSPF recomputation as satellites move, and neighbor state determines whether a next hop is usable); and transit packets are forwarded from the same per-namespace FIB.

The per-namespace FIBs are separate, but the path used to install routes into them is not.

\subsection{LEO Workload Characteristics}

LEO constellations differ from datacenter or ISP topologies in three ways that matter for emulation:

\textbf{(i) Scale.} Modern LEO constellations deploy thousands of satellites in coordinated orbital shells. Starlink Shell-1 comprises 1{,}584 satellites arranged in a 72$\times$22 grid; Kuiper Phase~1 will deploy 3{,}236. Emulating such topologies requires thousands of concurrent routing instances on a single host.

\textbf{(ii) High churn rate.} Unlike terrestrial networks where topology is largely static, LEO topologies are in constant motion. In a +Grid topology, intra-plane inter-satellite links are largely stable; the changes are concentrated in inter-plane links, which are torn down or re-established every few minutes as satellites enter the polar regions or cross the seam. Each topology change invalidates a large set of shortest paths simultaneously, triggering route recomputations across the entire network.

\textbf{(iii) Synchronized updates.} Topology changes often affect many satellites concurrently. For example, a plane handover event can simultaneously alter connectivity for dozens of satellites, creating synchronized route-update bursts that arrive at the shared kernel installation path in tight temporal clusters.

Prior emulation frameworks (e.g., Mininet~\cite{mininet}, Kathara~\cite{kathara}) were designed for smaller, more static topologies where route installations are rare events. Recent work (OpenSN~\cite{opensn}, StarryNet~\cite{starrynet}, SplitNN~\cite{splitnn}) has focused on optimizing the \emph{topology setup} phase---container creation, link provisioning, and initial configuration---which is a one-time cost paid before the emulation begins. However, the \emph{runtime} phase, where routing protocols continuously update forwarding state in response to topology changes, has received less attention. At LEO scale, this runtime installation path becomes the dominant bottleneck, as we demonstrate in Section~\ref{sec:rootcause}.

\begin{figure}[t]
\centering
\includegraphics[width=\columnwidth]{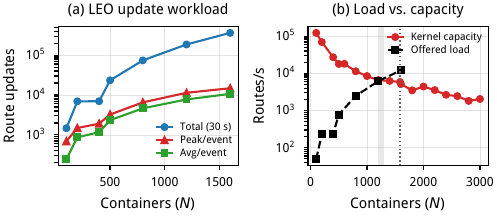}
\caption{(a) Route updates generated by topology events versus constellation size: total over 30\,s, and average and peak per event. (b) Kernel installation throughput and average offered load; shaded: crossing; dotted: Starlink Shell-1.}
\label{fig:leo-workload-kernel}
\end{figure}

\subsection{The Scalability Barrier}
\label{sec:barrier}

\textbf{We define sustainable capacity as the maximum long-run route-installation rate that does not lead to unbounded backlog.} This definition matters because short bursts are not the real problem. A system remains usable if it can absorb a burst and then drain the queue. It becomes unusable when pending route updates continue to accumulate even after the burst subsides.

Under this definition, the baseline kernel path exhibits a clear scalability barrier (Figure~\ref{fig:leo-workload-kernel}b). Kernel capacity falls from $1.2\times10^5$ routes/s at 100 containers to $6.6\times10^3$ at 1,200 and $5.2\times10^3$ at 1,600. Offered load moves the other way: the average update rate rises from 779 routes/s at 500 satellites to $6.2\times10^3$ at 1,200 and $1.2\times10^4$ at 1,600. At 1,200 containers the load already uses 94\% of the kernel's capacity, and at 1,600 it exceeds the capacity by a factor of 2.3. The crossing therefore lies at about 1,200--1,300 containers. Below it, the system absorbs bursts and drains its queue. Above it, backlog keeps growing and the data plane lags further and further behind the control plane.

Starlink Shell-1, at 1{,}584 satellites, is already past the barrier, and Kuiper Phase~1, at 3{,}236, is far beyond it. A $N^{2.35}$ fit of the measured load predicts about $5.4\times10^4$ routes/s at 3,000 satellites. The crossing thus falls inside the range of constellations researchers study today.

\subsection{Why Single-Machine Scalability Matters}

A single machine is the deployment most labs can afford and reproduce: there is no inter-machine network in the measurement path, so results are not confounded by cross-host latency, clock skew, or a synchronization protocol. Moving to multiple hosts requires tunneling the inter-node links, adds a synchronization layer, and makes results depend on the interconnect and the partitioning strategy. Raising the single-machine ceiling therefore keeps more experiments in this cleaner regime and delays the transition to distributed infrastructure.

\section{Root Cause Analysis}
\label{sec:rootcause}

\subsection{Methodology}

We profile large-scale containerized routing experiments with \texttt{perf} and flame graphs. The profiles show that, at scale, CPU time concentrates in synchronization-heavy kernel paths rather than in FRR route computation. The hot path follows a familiar chain: \texttt{RTM\_NEWROUTE} enters rtnetlink, acquires \texttt{rtnl\_mutex}, updates the FIB, and then triggers route notification handling through \texttt{netlink\_broadcast\_filtered}.

Profiling alone is not enough, because it shows where time is spent but not which component creates the scalability barrier. We therefore decompose route installation into three parts: namespace-local FIB update cost, global synchronization cost, and netlink notification cost. To isolate them, we use dedicated route injector tools and synthetic listeners rather than relying only on FRR behavior. This lets us vary one scaling dimension at a time and attribute throughput degradation to a specific mechanism.

\subsection{Isolating the Bottleneck}

\begin{figure*}[t]
\centering
\includegraphics[width=\textwidth]{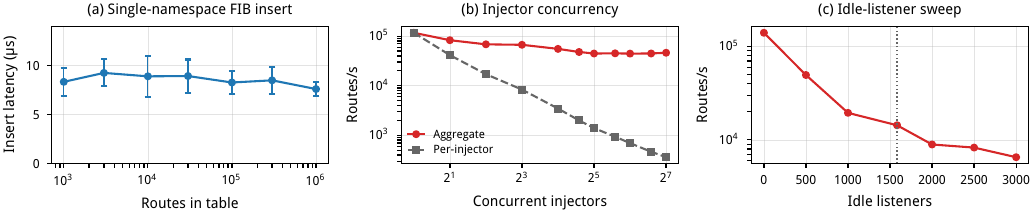}
\caption{Isolation experiments. (a) Per-route FIB insert latency in one namespace versus table size. (b) Aggregate and per-injector throughput versus concurrent injectors. (c) Throughput versus idle listeners in unrelated namespaces.}
\label{fig:route-injector-listener}
\end{figure*}

We first ask whether the FIB data structure inside one namespace is itself the bottleneck. In a single-injector, single-namespace experiment, we increase the number of installed routes and measure per-route installation latency. The result stays nearly flat as the local table grows from $10^3$ to $10^6$ routes (Figure~\ref{fig:route-injector-listener}a). This indicates that trie manipulation is not the dominant bottleneck. We therefore focus the remainder of the isolation study on experiments that vary shared synchronization and listener state.

We next isolate global synchronization by running multiple injectors concurrently while keeping the rest of the environment controlled. Aggregate throughput drops rapidly from the single-injector baseline, then settles into a broad plateau of $4.4$--$4.6\times10^4$ routes/s once 32--128 injectors contend at the same time. This confirms that \texttt{rtnl\_mutex} limits concurrency: adding injectors does not produce commensurate aggregate throughput gains, and per-injector throughput collapses. The plateau, however, also shows what the lock does \emph{not} explain: once writers are serialized, adding more of them leaves aggregate throughput roughly unchanged. Serialization caps throughput, but it does not by itself make the system slower as the emulation grows.

The key experiment varies the number of netlink listeners and namespaces while holding the number of active injectors fixed. We create listeners in idle namespaces that are not the target of any route updates. These listeners subscribe to route notifications via \texttt{NETLINK\_ROUTE} sockets but, due to namespace filtering, should never receive messages from the active injection workload. In an ideally isolated system, idle subscribers in separate namespaces should have no effect on route-installation throughput in the workload namespaces.

Instead, throughput drops sharply as listener count increases. Figure~\ref{fig:route-injector-listener}(c) shows this behavior directly. The kernel scans all $N$ listeners during each route update, but namespace matching rejects nearly all of them, producing almost zero actual deliveries. The system pays the full traversal cost inside the critical section for subscribers that are logically irrelevant to the workload.

\subsection{End-to-End Analysis}

\begin{figure}[t]
\centering
\includegraphics[width=\columnwidth]{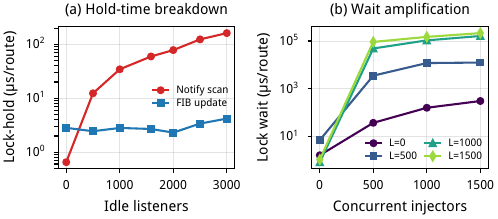}
\caption{(a) Per-route lock-hold time, split into FIB update and notification scan, versus idle listeners. (b) Per-route lock-acquire wait versus concurrent injectors, for four listener counts.}
\label{fig:timing-breakdown}
\end{figure}

Having identified the likely bottleneck, we return to the route-installation critical path and separate the per-route cost into lock-hold work and lock-acquire wait. Figure~\ref{fig:timing-breakdown} presents these two views together. Panel~(a) shows the two lock-hold components measured inside the serialized path while sweeping the number of idle listeners: namespace-local FIB update work and netlink notification scanning. As listeners increase, notification scanning grows much faster than FIB update work, revealing that listener-dependent notification cost is the dominant reason the critical section becomes longer. Panel~(b) then shows the resulting lock-acquire wait under increasing injector concurrency for four listener levels. Once the critical section is stretched by listener scanning, queueing amplifies that added hold time into a much larger wait penalty as more injectors contend simultaneously.

\subsection{Coupling Mechanism}

We can now state the mechanism precisely (red box in Figure~\ref{fig:baseline-path}). Network namespaces provide logical isolation of routing state, but the baseline route-installation path is implemented through shared kernel machinery. Every route update competes for the same \texttt{rtnl\_mutex}. Inside that serialized region, each update also incurs a scan whose cost depends on the total number of listeners and namespaces in the system. The consequence is hidden cross-container coupling: one container's control-plane activity slows down another's even when their routing state is logically separate.

The degradation with scale is therefore driven by the number of live subscribers, not by the number of concurrent writers: the mutex turns per-update cost into a system-wide cap, and the scan lowers that cap as routing daemons are added. If each route update pays an $O(N)$ scan cost and all $N$ containers generate updates concurrently, aggregate system cost per unit time approaches $O(N^2)$ in the regime where updates are densely packed.

\section{IsoRoute Design}
\label{sec:design}

\subsection{Design Requirements}
\label{sec:design-requirements}

\textbf{R1: Route installation isolated between nodes.} In the baseline, every route installation takes the global \texttt{rtnl\_mutex} and, inside it, scans the subscribers of all namespaces (\S\ref{sec:rootcause}), so how fast one node installs a route depends on how many nodes exist on the host. IsoRoute requires that installations at different nodes share no lock and notify no one else: the cost of installing a route at node $A$ must not depend on the total number of nodes, and updates at different nodes must be able to proceed in parallel on different cores. Both the forwarding tables (\S\ref{sec:design-tables}) and the agent that writes them (\S\ref{sec:design-install}) are designed around this requirement.

\textbf{R2: Deployability, with almost no change to how an emulation is built and operated.} IsoRoute must be usable on existing container-emulation platforms. The host kernel is neither modified nor replaced, because emulation experiments usually run on shared lab servers whose users cannot patch the kernel or load kernel modules. FRR may change only minimally, otherwise every FRR upgrade requires a new port. Topology scripts and experimental workflows should also stay essentially unchanged. Figure~\ref{fig:changes} lists all of IsoRoute's changes: one line in \texttt{zebra}, two lines of FPM configuration, and, in each container, one default route and two policy routing rules; the minimum kernel version is stated in \S\ref{sec:implementation}.

\textbf{R3: Emulation invariance, with forwarding behavior identical to the original hop-by-hop kernel forwarding.} After the forwarding plane is replaced, the emulated network must still be the same network. Every node still computes its own OSPF routes and forwards hop by hop from its own table. Routes must not be computed centrally and pushed down, and the network structure must not be changed to reduce load, for example by splitting a constellation into multiple OSPF areas~\cite{opensn}. Packet-level behavior must also match the kernel: the TTL decreases at each hop, TTL expiry returns an ICMP error, link delays still apply, tools such as \texttt{ping} and \texttt{traceroute} keep working inside containers, and packet capture tools such as Wireshark still observe traffic on the emulated links. The last requirement rules out designs that forward entirely outside the kernel's virtual devices. Two differences are visible: OSPF routes no longer appear in a container's \texttt{ip route} output (\S\ref{sec:discussion-questions}), and an application inside a container does not immediately receive an error when sending to a temporarily unreachable address (\S\ref{sec:design-forward}).

\textbf{Alternative designs.} Satisfying any two of the three requirements is straightforward; satisfying all three is not. Table~\ref{tab:alternatives} lists several obvious alternatives, each of which violates at least one.

\begin{table}[t]
\centering
\scriptsize
\setlength{\tabcolsep}{3pt}
\begin{tabularx}{\columnwidth}{@{}>{\raggedright\arraybackslash}p{0.22\columnwidth}ccc>{\raggedright\arraybackslash}X@{}}
\toprule
Approach & R1 & R2 & R3 & Problem \\
\midrule
Modify the kernel & \checkmark & $\times$ & \checkmark & Requires replacing the host kernel; upstream per-netns RTNL locking~\cite{pernetns-rtnl} does not remove the subscriber scan \\
DPDK/VPP userspace forwarding & \checkmark & $\times$ & $\times$ & Needs hugepages and dedicated polling cores; veth must become memif or TAP; packets bypass the kernel, so \texttt{netem} delay, ICMP, ARP and local traffic all have to be reimplemented in userspace \\
Intercept Netlink~\cite{lfri} & \checkmark & $\times$ & partial & The interposer must re-implement Netlink acknowledgments, dumps and notifications; its forwarder implements link- and network-layer processing itself, and parity with the kernel is unverified \\
One eBPF table for all nodes & $\times$ & \checkmark & \checkmark & The kernel protects an entire \texttt{LPM\_TRIE} with one lock, re-serializing all nodes' updates \\
One agent and eBPF program per container & \checkmark & $\times$ & \checkmark & Loading eBPF inside containers needs privileges such as \texttt{CAP\_BPF} and modified images; process count and memory grow linearly with the node count \\
Full router in eBPF & \checkmark & \checkmark & $\times$ & ICMP, ARP and local traffic must be implemented from scratch and can easily diverge from the kernel \\
IsoRoute & \checkmark & \checkmark & \checkmark & \\
\bottomrule
\end{tabularx}
\caption{Alternative designs and the design requirements.}
\label{tab:alternatives}
\end{table}

\subsection{Overview}
\label{sec:design-overview}

\begin{figure}[t]
\centering
\includegraphics[width=\columnwidth]{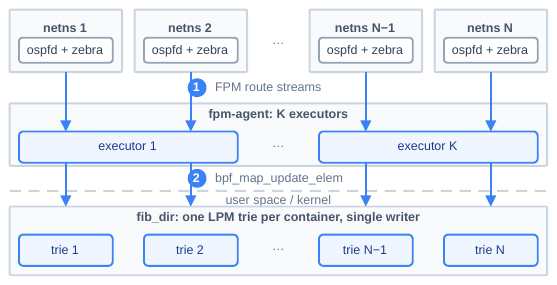}
\caption{IsoRoute route installation. \textbf{(1)} \texttt{zebra} exports routes over FPM. \textbf{(2)} The \texttt{fpm-agent} shards containers across $K$ executors; each executor writes into its containers' own LPM tries, so every trie has a single writer.}
\label{fig:isoroute-arch}
\end{figure}

IsoRoute organizes route installation as three stages (Figures~\ref{fig:isoroute-arch} and~\ref{fig:isoroute-dataplane}).

\begin{enumerate}[leftmargin=1.4em,itemsep=0pt,topsep=2pt]
\item \textbf{Export.} \texttt{ospfd} hands computed routes to \texttt{zebra} over ZAPI as before. Instead of writing them into the kernel FIB, \texttt{zebra} streams them over FPM to a host-side \texttt{fpm-agent}.
\item \textbf{Install.} The agent attributes each FPM stream to an emulated node and writes every route into that node's own longest-prefix-match (LPM) table, an eBPF map.
\item \textbf{Forward.} A single tc program, shared by all nodes, looks up each IPv4 packet in the table of the node that received or sent it and redirects it to the next hop. Everything the program does not handle continues through the kernel.
\end{enumerate}

\textbf{Walkthrough.} Suppose an inter-satellite link of satellite $A$ fails. \texttt{ospfd} in $A$ floods the change and every affected node recomputes its routes, exactly as in the baseline. In $A$, \texttt{zebra} selects the new routes and sends, for each changed prefix, one FPM message to the agent. The agent identifies the stream as $A$'s, places the updates in $A$'s queue, and applies them in order to $A$'s table with one map operation each; updates of other nodes proceed in parallel on other cores. The next packet that $A$ forwards towards an affected destination is redirected to the new next hop. At no point does the update take a lock shared with other nodes or notify other nodes' sockets (R1).

The following sections follow the three requirements: \S\ref{sec:design-tables} and \S\ref{sec:design-install} realize node isolation at the forwarding-table and installation-path layers (R1), and \S\ref{sec:design-install} also shows how a single FRR change keeps routes out of the kernel (R2); \S\ref{sec:design-forward} shows how hop-by-hop forwarding behavior is preserved after the forwarding plane is replaced (R3).

\textbf{What changes.} Figure~\ref{fig:changes} lists every change IsoRoute makes to the baseline environment: one changed line in \texttt{zebra}'s Netlink dataplane, two lines of FPM configuration, and, in each container, one default route and two policy routing rules. Everything else is unmodified.

\begin{figure}[t]
\begin{lstlisting}[style=routercode,numbers=none,basicstyle=\ttfamily\scriptsize]
# (1) zebra/kernel_netlink.c, nl_put_msg()  [1-line patch]
 case DPLANE_OP_ROUTE_INSTALL:
 case DPLANE_OP_ROUTE_UPDATE:
 case DPLANE_OP_ROUTE_DELETE:
-    return netlink_put_route_update_msg(bth, ctx);
+    return FRR_NETLINK_SUCCESS;

# (2) zebra configuration  [FPM export]
+ fpm address <agent-ip>
+ no fpm use-next-hop-groups

# (3) each container  [local traffic, Sec. 4.5]
+ ip rule add iif lo lookup main pref 100
+ ip rule add iif lo lookup 100  pref 101
+ ip route add default dev dummy0 src <loopback-ip> table 100
\end{lstlisting}
\caption{Every change IsoRoute makes to a baseline emulation node. Lines marked \texttt{+} are added and \texttt{-} removed; the kernel and all FRR daemons other than this \texttt{zebra} function are unmodified.}
\label{fig:changes}
\end{figure}

\subsection{Per-Node Forwarding Tables}
\label{sec:design-tables}

R1 first requires a forwarding table that belongs to one node alone, so that updating it does not involve other nodes. The simplest layout is one shared LPM table keyed by (node, prefix), but the kernel protects an entire \texttt{LPM\_TRIE} with a single lock, so a shared table would move the global serialization of \S\ref{sec:rootcause} into eBPF. At the other extreme, loading one program with its own table per node makes the number of programs grow with the topology. A further difficulty is that a transit packet carries no socket, so the program cannot tell which node it belongs to. IsoRoute identifies nodes by network-namespace cookie, lets all nodes share one program, and reaches their tables through two levels of maps.

\textbf{Node identity: the network-namespace cookie.} Identifying a node by interface name, interface index or container name would tie IsoRoute to the topology builder's naming conventions, which can also break when containers are recreated. IsoRoute instead uses the network-namespace cookie, assigned by the kernel and unique on the host. The forwarding program converts its context with \texttt{bpf\_cast\_to\_kern\_ctx} and reads, through CO-RE, the cookie of the namespace owning the \emph{device} the packet is on (\texttt{skb->dev->nd\_net->net\_cookie}). The agent enters the container's namespace and reads \texttt{SO\_NETNS\_COOKIE} for the same value. Both sides therefore use one identity without any naming convention.

\textbf{Table layout: one program, one table per node.} IsoRoute loads a single program and reaches per-node tables through two maps (Appendix~\ref{sec:ebpf-code}). \texttt{cookie2rid} is a hash map from a node's cookie to a small integer table identifier (\texttt{rid}); \texttt{fib\_dir} is a \texttt{HASH\_OF\_MAPS} from \texttt{rid} to that node's inner \texttt{LPM\_TRIE}. Each inner entry maps a destination prefix to an egress interface index and a gateway. Growing the topology adds map entries, not programs, at the cost of one extra lookup per packet. The \texttt{rid} indirection lets the agent replace a node's whole table with one map update (\S\ref{sec:design-install}). Inner tables are created when a node's first route arrives and allocated with \texttt{BPF\_F\_NO\_PREALLOC}, so memory grows with the routes actually installed.

This layout satisfies R1 at the forwarding-table layer: the kernel protects each LPM trie with its own lock, so an update to node $A$ touches only $A$'s trie, notifies no one, and costs nothing that depends on the number of other nodes.

\subsection{The Bypassed Installation Path}
\label{sec:design-install}

The path from \texttt{zebra} to per-node tables must satisfy two requirements at once: it must avoid the kernel installation path without changing the kernel and with only a small change to FRR (R2), and it must contain no serialization point shared by all nodes (R1). Two difficulties arise. First, \texttt{zebra} must believe that a route has been installed so that its state machine advances, while its link, address and neighbor interactions with the kernel must be preserved. Second, routes have to converge on some process that writes the tables: one agent per container requires modified images, eBPF privileges inside containers and resources that grow with the node count, whereas one agent for all nodes could become a new serialization point. IsoRoute uses FRR's own FPM interface to export routes to one agent on the host, which shards the writes by node.

\textbf{Exporting routes over FPM.} FRR is widely regarded as one of the state-of-the-art open-source routing suites, so we scale emulation around it rather than replacing it; \texttt{zebra} already installs routes incrementally~\cite{frr-src}, and IsoRoute takes this as a given---redundant-installation filtering of the kind studied in LFRI~\cite{lfri} is therefore not our optimization target. Interposing on \texttt{zebra}'s Netlink traffic would also capture routes, but the interposer must then re-implement Netlink semantics for everything it diverts, including acknowledgments, dumps and notifications. FRR already provides FPM, which \texttt{zebra} uses to feed forwarding planes outside the kernel, such as switch ASICs in SONiC~\cite{sonic}. FPM streams every install, replace and withdrawal as a standard Netlink route message over TCP, and leaves \texttt{zebra}'s other Netlink use---link, address and neighbor state---untouched. We also configure \texttt{no fpm use-next-hop-groups}, so each message carries its gateway and output interface directly. The agent then needs no next-hop-group state, and each message maps to exactly one map operation.

\textbf{A one-line patch stops kernel installation.} Enabling FPM alone is not enough. \texttt{zebra} still writes the kernel FIB~\cite{frr-fpm}, and in its dataplane FPM sits after the kernel provider, so every route passes through the kernel installation path before it is exported. We change one line in \texttt{zebra}'s \texttt{nl\_put\_msg()}: for route install, update and delete operations it returns \texttt{FRR\_NETLINK\_SUCCESS} instead of calling \texttt{netlink\_put\_route\_update\_msg()}, which encodes the request for the kernel (Figure~\ref{fig:changes}). The update then continues to FPM as if the kernel had accepted it. This is the only change to FRR.

\textbf{Validation experiment.} We checked this with a standalone \texttt{zebra} into which \texttt{sharpd} injected 1,000 routes: with the patch, \texttt{zebra} issued no route request to the kernel, counted all 1,000 routes as installed, and exported all of them over FPM.

\textbf{Attributing streams to nodes.} An agent could infer a stream's owner from a naming convention or from configuration pushed by the topology builder, but both break when containers are recreated. Instead, each \texttt{zebra} connects to the agent over the management network, and the agent maps the connection's source address to a container through the container runtime, then to that container's network-namespace cookie (\S\ref{sec:design-tables}). Container metadata can appear later than the FPM connection, so the lookup is retried every 500\,ms for up to two minutes.

\textbf{Sharded executors.} The simplest agent gives every FPM connection its own goroutine that decodes the stream and writes the map directly. This preserves order and runs nodes in parallel, but with thousands of connections it lets thousands of goroutines issue map updates at once, with no control over which core handles which node.

IsoRoute separates decoding from installation. One reader per connection decodes the stream; installation is done by a fixed set of executor threads, one per CPU by default, pinned and interleaved across sockets. Each node is assigned to one executor by hashing its cookie, and each node has its own FIFO mailbox, so its updates are applied in order while nodes on different executors never wait for each other. Readers hand updates to executors through a lock-free stack that the executor drains in batches, at most 64 updates from one mailbox per round, so a busy node cannot starve others. The hot path takes no lock; locks are taken only when a node's first route arrives and its table identifier and table are created. Nodes on the same executor share only CPU time: the number of executors equals the number of CPUs and the cost of one map write does not grow with $N$, so R1 still holds. Writing the table directly from each connection remains available as a configuration option.

\textbf{Resynchronization.} FPM uses replace semantics, and \texttt{zebra} resends its complete table when an FPM connection is re-established. Deleting the old entries one by one before applying the resent table would leave a window in which the node forwards with a mix of old and new routes. IsoRoute replaces the table instead: when the first connection of a node arrives, the agent creates an empty LPM trie, swaps it into the node's slot in \texttt{fib\_dir} with one map update (\S\ref{sec:design-tables}), and discards updates of the previous connection that are still queued, so they cannot overwrite the new state. A one-second window suppresses duplicate resets when several connections of the same node arrive together. The slot always points to a valid table during the swap, so the forwarding program never finds it missing.

\subsection{Bypassing Only the FIB Lookup}
\label{sec:design-forward}

\begin{figure}[t]
\centering
\includegraphics[width=\columnwidth]{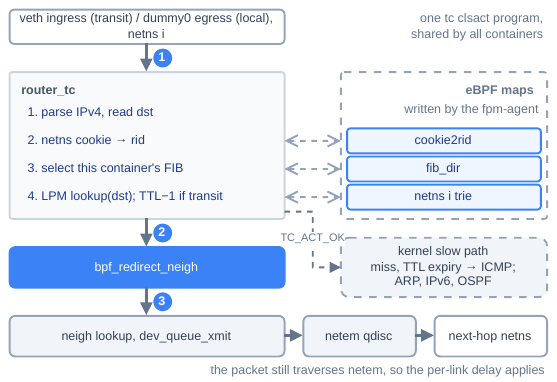}
\caption{IsoRoute dataplane. A shared tc program selects the node's trie by netns cookie and redirects hits via \texttt{bpf\_redirect\_neigh}; all other packets are handled by the kernel as usual.}
\label{fig:isoroute-dataplane}
\end{figure}

R3 requires forwarding behavior identical to hop-by-hop kernel forwarding. Once OSPF routes leave the kernel FIB, the difficulties follow: packets from applications inside a container fail the kernel route lookup, and ICMP errors, neighbor resolution and \texttt{netem} link delay all depend on the kernel forwarding path. Re-implementing these in eBPF or DPDK duplicates a large amount of work and easily introduces small differences from the kernel. IsoRoute replaces exactly one step in a node's packet processing: the FIB lookup during forwarding. Reception, local delivery, ICMP generation, neighbor resolution, queueing and transmission remain with the kernel; the eBPF program only looks up a route, decrements the TTL and redirects the packet to the next hop. It is attached at two points: ingress of every inter-satellite link, to intercept transit packets, and egress of \texttt{dummy0}, to intercept packets originated by the node (Appendix~\ref{sec:ebpf-code}, Figure~\ref{fig:isoroute-dataplane}). The management interface used by FPM is left unattached.

\textbf{Transit packets: table lookup at link ingress.} When a packet arrives on an inter-satellite link, the program first passes non-IPv4, multicast and OSPF packets through, then finds the node's table by the netns cookie of the device the packet is on (\S\ref{sec:design-tables}) and looks up the destination. On a hit, it checks and decrements the TTL, updates the checksum incrementally, and calls \texttt{bpf\_redirect\_neigh()} to send the packet to the next hop, skipping the kernel's FIB lookup. Misses and packets about to expire are passed unchanged to the kernel: packets addressed to this node are delivered to the local stack; on TTL expiry the kernel generates ICMP Time Exceeded, so \texttt{traceroute} works; and when no route to the destination exists yet---for example, after OSPF has withdrawn an old route but before the new one is installed---the kernel drops the packet and returns ICMP Destination Unreachable, as in the baseline.

\textbf{Local packets: entering eBPF through \texttt{dummy0}.} OSPF routes exist only in the eBPF tables; the container's kernel FIB holds only connected and local routes. Without further changes, a packet from an application inside the container to a remote node would fail the kernel route lookup (\texttt{Network is unreachable}) and never reach the eBPF program. IsoRoute gives each container a \texttt{dummy0} device and places a route \texttt{default dev dummy0} in a separate routing table, with the node's loopback address as preferred source.

Two policy routing rules (Figure~\ref{fig:changes}) make only locally originated packets (\texttt{iif lo}) consult that table, and consult the main table first, so packets to directly connected neighbors and to the management network (including the FPM connection) still use their original routes. Applications such as \texttt{ping} and \texttt{iperf3} therefore need no modification: the kernel hands their packets to \texttt{dummy0}, and the program at the \texttt{dummy0} egress hook looks them up and redirects them to the right inter-satellite link. Like a host, a node does not decrement the TTL of packets it originates; the program in Appendix~\ref{sec:ebpf-code} distinguishes the two cases with \texttt{ingress\_ifindex}. ICMP packets generated by the kernel are locally originated traffic too, and leave through \texttt{dummy0}.

Restricting the default route to local packets is necessary for R3. If transit packets could match it, the kernel would forward packets with no route to \texttt{dummy0}, where they would be dropped, instead of returning ICMP Destination Unreachable as in the baseline.

\textbf{Everything else stays with the kernel.} Neighbor resolution is delegated to the kernel by \texttt{bpf\_redirect\_neigh()}, so the agent keeps no ARP state. Redirected packets leave through the normal transmit path and traverse the egress qdisc, so \texttt{netem} delays configured on veth ends stay in effect (\S\ref{sec:dataplane}). Because packets still enter and leave the containers through their veth devices, a packet capture on a link (e.g., with Wireshark) sees the same frames as in the baseline; this compatibility is another reason not to bypass the virtual devices entirely, as a userspace forwarder would. OSPF, multicast, ARP and IPv6 packets are passed to the kernel unchanged.

\textbf{Difference from kernel behavior.} One difference from the baseline remains. When an application inside a container sends to a temporarily unreachable address, the baseline kernel fails the route lookup immediately and the application receives \texttt{Network is unreachable}; in IsoRoute the packet first follows the default route into \texttt{dummy0}, is dropped on a table miss, and the application has to wait for a timeout. Transit packets are unaffected.

\textbf{Why tc and not XDP.} XDP would save some per-packet work but cannot be used here: \texttt{bpf\_redirect\_neigh()} is available only to tc programs; identifying the node requires reading the device's namespace from the \texttt{sk\_buff}; and locally originated packets must be intercepted at an egress hook, which XDP does not have.

\section{Implementation}
\label{sec:implementation}

\textbf{Agent.} The \texttt{fpm-agent} is written in Go (1.24) and uses \texttt{cilium/ebpf} to load the program and manage maps and the Docker API to resolve containers. Its core, the FPM server, decoder and executor pipeline (about 4,200 lines) and the map manager (about 3,600 lines), is about 7,800 lines; the command-line front end adds configuration handling and the measurement hooks used in \S\ref{sec:evaluation}. Executor count, mailbox batch size, ingress queue capacity and inner-table size are configurable; the defaults are one executor per CPU, 64 updates per mailbox round, 512 queued events per executor and 200,000 entries per node table. To keep per-update cost low, map writes on the hot path issue the \texttt{bpf} system call directly, and the agent caches each node's table handle and prepared map values.

\textbf{eBPF program.} The forwarding program is 272 lines of C compiled with clang and loaded through CO-RE; it is attached with tc \texttt{clsact} hooks by the topology scripts. It relies on \texttt{bpf\_redirect\_neigh()} (available since Linux 5.10) and on the \texttt{bpf\_cast\_to\_kern\_ctx} kfunc (added in Linux 6.2), so Linux 6.2 or newer is required; we run it on Linux 6.18.4.

\textbf{Control plane and topology.} We use FRR 9.1 with \texttt{zebra} loaded with the \texttt{dplane\_fpm\_nl} module and the patch of Figure~\ref{fig:changes}. Shell scripts build the constellation (containers, veth pairs, \texttt{netem} delays, \texttt{dummy0}, the default route and the policy routing rules), generate per-node FRR configurations, enable FPM, and attach the program to link and \texttt{dummy0} interfaces in parallel.

\section{Evaluation}
\label{sec:evaluation}

\subsection{Experimental Setup}
\label{sec:eval-setup}

We compare IsoRoute against the standard FRR+Linux stack, in which \texttt{zebra} installs routes into each container's kernel FIB through Netlink. Both systems use the same topology generator, the same routing software and the same constellation scales; they differ only in how forwarding state is installed. The baseline is unmodified FRR, which already installs only routes whose selected next hops change, so all improvements below are measured on top of incremental installation.

\begin{table}[t]
\centering
\small
\resizebox{\columnwidth}{!}{%
\begin{tabular}{ll}
\toprule
Component & Configuration \\
\midrule
Host & 36 cores @ 2.19\,GHz, 256\,GB RAM, VMware ESXi \\
Guest VM & 36 vCPUs, 180\,GB RAM \\
OS / kernel & Debian 12, Linux 6.18.4 (xanmod) \\
Routing software & FRR 9.1 (\texttt{ospfd}, \texttt{zebra}, \texttt{sharpd}) \\
Containers & Docker, one network namespace per node \\
\midrule
\multicolumn{2}{l}{\textit{IsoRoute}} \\
\texttt{fpm-agent} & Go; sharded executors (\S\ref{sec:design-install}) \\
eBPF attachment & tc ingress on links, egress on \texttt{dummy0} \\
\bottomrule
\end{tabular}}
\caption{Testbed configuration.}
\label{tab:testbed}
\end{table}

Table~\ref{tab:testbed} lists the testbed. All experiments run on a single machine so that results are not affected by distributed-system effects. The LEO workload comes from Starlink-like grids in which every satellite runs \texttt{ospfd} and \texttt{zebra} in its own container; link changes trigger OSPF recomputation and bursts of route installations. For microbenchmarks we inject routes with FRR's \texttt{sharpd} to isolate the installation path. Kernel capacity is measured at 18 scales from 100 to 3,000 containers with one run per scale; IsoRoute capacity is measured at six scales from 500 to 3,000 with 10 runs each (7 at 3,000). IsoRoute points in Figures~\ref{fig:scale-sweep} and~\ref{fig:multicore-scaling} are means of the runs, and the error bands are 95\% confidence intervals.

\S\ref{sec:eval-barrier} and \S\ref{sec:eval-multicore} test node isolation (R1): whether installation capacity falls with scale and whether it scales with core count. \S\ref{sec:eval-e2e} reports end-to-end OSPF convergence time. \S\ref{sec:dataplane} tests emulation invariance (R3) and reports the agent's resource cost. The evidence for R2 is the size of the change (Figure~\ref{fig:changes}): no kernel change, one line in FRR, and only the \texttt{dummy0} device and its routing configuration added to the topology scripts.

\subsection{Breaking the Scalability Barrier}
\label{sec:eval-barrier}

Figure~\ref{fig:scale-sweep} compares installation capacity with the offered load. Kernel capacity falls steadily with scale, from $1.2\times10^5$ routes/s at 100 containers to $2.0\times10^3$ at 3,000, and drops below the offered load at about 1,200--1,300 containers. IsoRoute capacity stays between $3.0\times10^5$ and $3.5\times10^5$ routes/s across the whole range from 500 to 3,000 containers. At 3,000 containers it is about 170 times the kernel's capacity and about six times the extrapolated offered load. Table~\ref{tab:barrier-summary} summarizes what this means for the two constellations of interest.

\begin{table}[t]
\centering
\small
\resizebox{\columnwidth}{!}{%
\begin{tabular}{lccc}
\toprule
Approach & Crossing & Starlink (1,584) & Kuiper (3,236) \\
\midrule
Kernel & $\approx$1,200--1,300 & No & No \\
IsoRoute & none up to 3,000 & Yes & Yes$^\ast$ \\
\bottomrule
\end{tabular}}
\caption{Whether installation capacity exceeds the average offered load. $^\ast$IsoRoute capacity is measured up to 3,000 containers; Kuiper's offered load is extrapolated.}
\label{tab:barrier-summary}
\end{table}

\subsection{Throughput and Multi-Core Scalability}
\label{sec:eval-multicore}

\begin{figure}[t]
\centering
\includegraphics[width=\columnwidth]{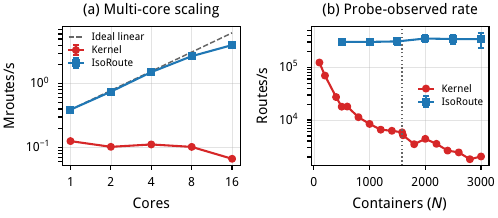}
\caption{(a) Multi-core scaling; the dashed line is ideal linear scaling from the 1-core IsoRoute point. (b) Install rate observed with bpftrace at the route-installation hooks, across scale.}
\label{fig:multicore-scaling}
\end{figure}

If IsoRoute removes the shared serialization point, extra cores should translate into extra installation throughput. Figure~\ref{fig:multicore-scaling}(a) tests this with 1, 2, 4, 8 and 16 cores. Kernel throughput stays between $0.10$ and $0.13$ Mroutes/s from 1 to 8 cores and drops to $0.066$ Mroutes/s at 16 cores. IsoRoute rises from $0.391$ Mroutes/s on one core to $2.695$ Mroutes/s on 8 cores and $4.023$ Mroutes/s on 16 cores.

Dividing those numbers by the core count gives the parallel efficiency: the kernel path reaches 0.81 times its 1-core throughput at 2 cores and 0.53 times at 16 cores, while IsoRoute reaches 0.96, 0.97 and 0.86 at 2, 4 and 8 cores and 0.64 at 16 cores. Updates of different nodes no longer serialize, although some efficiency is lost at the highest core count.

Figure~\ref{fig:multicore-scaling}(b) shows the install rate observed with bpftrace at the route-installation hooks of both systems, across scale. The observed rate of the kernel path collapses as $N$ grows, as expected from the per-update subscriber scan; IsoRoute stays high and nearly flat, because a map update touches only the owning node's table and notifies no one.

The two throughput figures are not directly comparable: \S\ref{sec:eval-barrier} reports sustainable installation capacity measured across emulation scales, whereas Figure~\ref{fig:multicore-scaling}(a) sweeps the number of executor cores to measure the peak rate of the installation path alone. We report them separately as a scale test and a core-scaling test.

\subsection{End-to-End OSPF Convergence}
\label{sec:eval-e2e}

Capacity numbers matter only if they change what an experiment observes. We therefore run OSPF on a Starlink Shell-1 topology of 1,584 nodes ($72\times22$ grid), with a Hello interval of 10\,s, a Dead interval of 40\,s and 10\,ms inter-satellite links. Using kernel tracing, we record the time at which each route becomes usable in the active forwarding plane. Data-plane convergence starts at the first route installation on any node and ends when no installation is observed for 5\,s; this window is the period during which forwarding state across the emulated network is inconsistent with the control plane.

\begin{figure}[t]
\centering
\includegraphics[width=0.92\columnwidth]{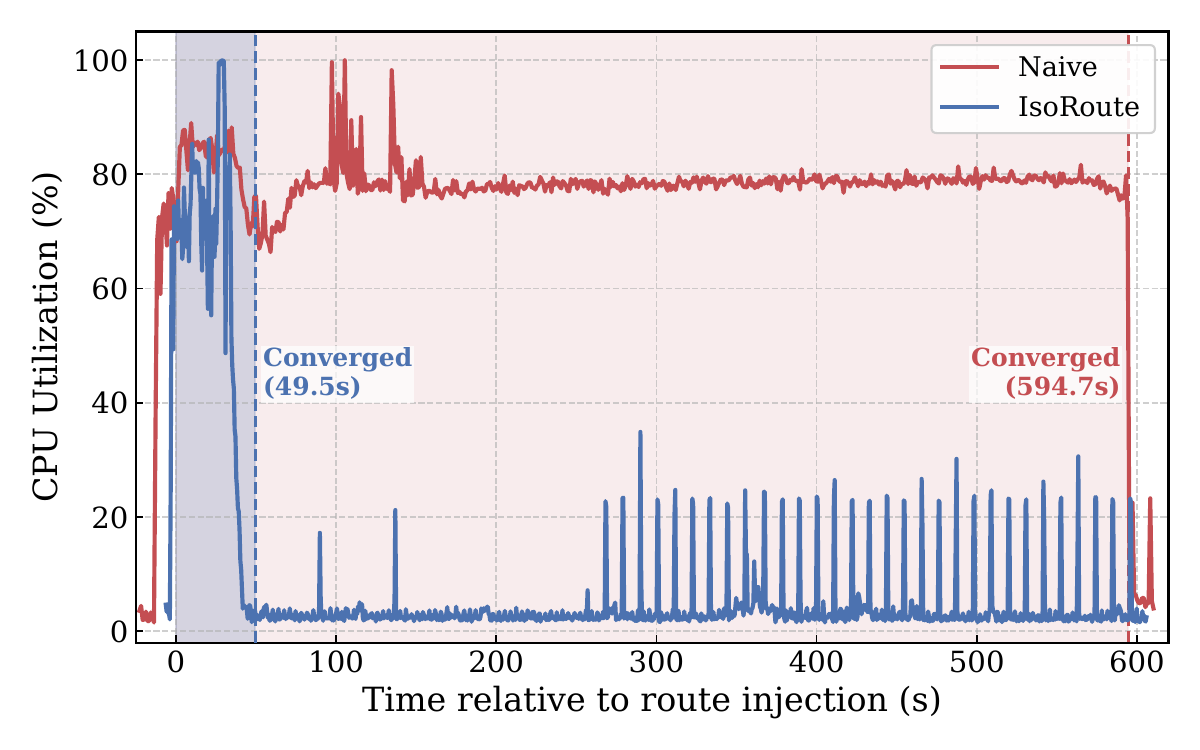}\\[-1pt]
{\footnotesize (a) Host CPU utilization}\\[3pt]
\includegraphics[width=0.80\columnwidth]{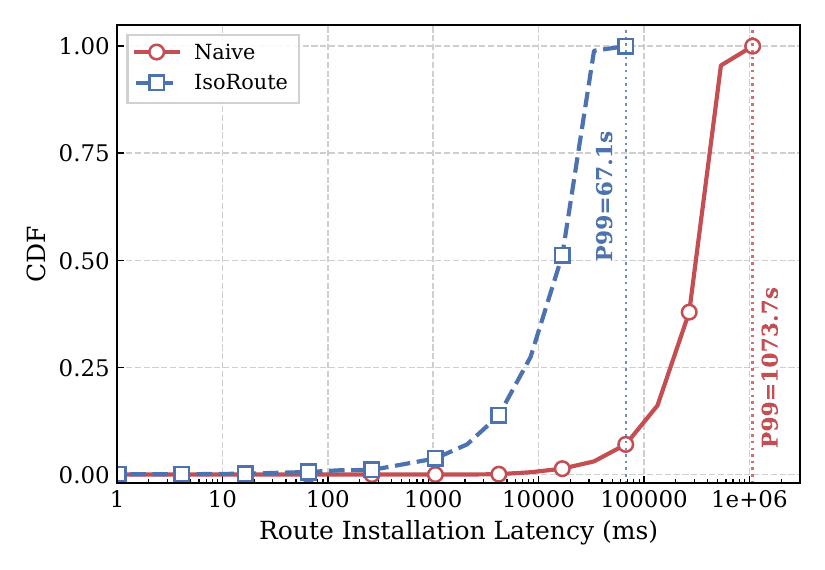}\\[-1pt]
{\footnotesize (b) Route-installation completion}
\caption{OSPF convergence on a 1,584-node Starlink Shell-1 topology (``Naive'' denotes the kernel baseline). (a) Host CPU; dashed: data-plane convergence. (b) CDF of per-route installation completion time.}
\label{fig:e2e}
\end{figure}

Figure~\ref{fig:e2e} shows the result. With the kernel path, the data plane needs 594.7\,s to converge, and host CPU utilization stays near 80\% for the whole period: route installations queue on the global mutex, and each pays a scan over the subscribers of all 1,584 nodes. IsoRoute converges in 49.5\,s, 12.0 times faster, after which CPU utilization falls to a few percent. The distribution of per-route completion times shows the same effect: the kernel path has a long tail (P99 1,073.3\,s), and IsoRoute removes most of it (P99 67.1\,s). Figure~\ref{fig:e2e}(b) comes from a separate run with additional tracing, so its absolute times can exceed the convergence time in (a). For an emulation, this difference decides whether the network under study is observed in its converged state or in a transient that exists only because of the emulator.

\subsection{Forwarding Correctness and Overhead}
\label{sec:dataplane}

IsoRoute replaces the kernel forwarding path, so we must show that it does not change what is being emulated.

\textbf{Forwarding-table equivalence.} We compare, entry by entry, each node's eBPF table with the routes that FRR has selected for installation (\texttt{show ip route json}): for every prefix, the egress interface and gateway must match. On a converged $6\times6$ grid (36 nodes), all 1,260 OSPF routes selected by FRR (35 per node) are present in the eBPF tables with the same egress interface and gateway, giving 0 mismatching entries out of 1,260. The tables additionally hold 144 entries, four per node, which are the subnets of the node's own links and appear in FRR as connected routes.

\begin{figure}[t]
\centering
\includegraphics[width=\columnwidth]{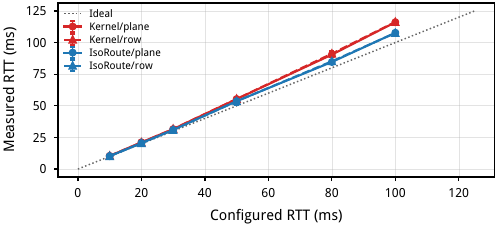}
\caption{Measured versus configured RTT on a $72\times22$ topology with 5\,ms links, for paths within an orbital plane and across planes. Each point aggregates 10 repeats of 100 pings at 0.1\,s intervals.}
\label{fig:dataplane-delay}
\end{figure}

\textbf{Link delay.} Emulated link delays are implemented as \texttt{netem} qdiscs on the veth ends. Packets redirected by the tc program leave through \texttt{dev\_queue\_xmit} and therefore still traverse these qdiscs. Figure~\ref{fig:dataplane-delay} compares measured and configured RTT on 1- to 10-hop paths of a $72\times22$ topology with 5\,ms links. Both systems apply the configured delays; at a configured 100\,ms RTT, IsoRoute measures 107.6--107.7\,ms and the kernel path 116.1--116.3\,ms, so IsoRoute adds less forwarding delay per hop than the kernel path.

\textbf{Applications inside containers.} On the same grid we run unmodified tools in node S1\_1 towards S4\_4, six hops away. \texttt{ip route get} resolves the destination to \texttt{dev dummy0}, so locally originated traffic enters the eBPF path through the egress hook. \texttt{ping} receives all 20 replies. \texttt{traceroute} lists the six routers of the path computed by OSPF, one per hop and in order, and ends at the destination. At an intermediate node, the kernel counter \texttt{IcmpOutTimeExcds} increases by one during the traceroute, i.e., the Time Exceeded message is generated by the kernel, while \texttt{IpForwDatagrams} stays at zero throughout, i.e., no transit packet is forwarded by the kernel IP stack.

\textbf{Agent cost.} On the 1,584-node topology, the idle agent uses about 1.9\,GB of resident memory, with one reader per FPM connection (1,584) and 16 executors.

\textbf{Feature coverage.} Table~\ref{tab:feature-coverage} lists the router behaviors that IsoRoute provides and which path provides them.

\begin{table}[t]
\centering
\small
\resizebox{\columnwidth}{!}{%
\begin{tabular}{lcl}
\toprule
Feature & IsoRoute & How \\
\midrule
IPv4 unicast forwarding & \checkmark & eBPF program \\
TTL decrement & \checkmark & eBPF, incremental checksum \\
Locally originated traffic & \checkmark & default route to \texttt{dummy0}, local packets only \\
ICMP Destination Unreachable & \checkmark & kernel (transit packets) \\
ICMP errors (e.g., Time Exceeded) & \checkmark & kernel \\
Neighbor resolution & \checkmark & kernel, \texttt{bpf\_redirect\_neigh} \\
\texttt{netem} link delay & \checkmark & egress qdisc preserved \\
Packet capture on links & \checkmark & veth devices retained in the path \\
\bottomrule
\end{tabular}}
\caption{Router behaviors in IsoRoute and the path that provides each.}
\label{tab:feature-coverage}
\end{table}

\section{Discussion}
\label{sec:discussion}

\subsection{Design Questions}
\label{sec:discussion-questions}

\textbf{Is IsoRoute still needed after upstream kernel improvements?} Upstream work on per-namespace RTNL locking~\cite{pernetns-rtnl} may reduce serialization in future kernels, but the subscriber scan in the notification path remains (Table~\ref{tab:alternatives}). Even if both were eventually removed, IsoRoute depends on no kernel modification and can keep running on existing distribution kernels.

\textbf{Does the agent become the new bottleneck?} The agent is a single process serving all nodes, so it could in principle reintroduce the serialization IsoRoute removes. Two measurements show that it does not. Installation throughput stays at $3.0$--$3.5\times10^5$ routes/s from 500 to 3,000 containers, i.e., it does not decline with the number of nodes (\S\ref{sec:eval-barrier}), and it scales to $4.0\times10^6$ routes/s on 16 cores (\S\ref{sec:eval-multicore}). Structurally, nodes share only the executor they hash to, and the hot path takes no lock (\S\ref{sec:design-install}).

\textbf{Does bypassing the kernel change what is emulated?} The control plane, its messages and its view of links and neighbors are the same as in the baseline (R3). Converged forwarding tables match the routes FRR selected (\S\ref{sec:dataplane}), configured link delays are preserved, and ICMP errors and locally originated traffic are still produced by the kernel as before. One visible difference remains: because OSPF routes live in eBPF tables, \texttt{ip route} inside a container shows only connected routes, and the default route to \texttt{dummy0} sits in a separate routing table. Researchers working with FRR already read routing state through \texttt{vtysh}, which shows the same selected routes as before, and applications inside containers do not depend on the kernel table because their traffic is forwarded through \texttt{dummy0}.

\subsection{Availability}
\label{sec:availability}

The \texttt{fpm-agent}, the eBPF programs, the topology scripts and the measurement scripts used in this paper are released at \url{https://github.com/akaixinguo/IsoRoute} and archived under the Zenodo DOI~\mbox{\href{https://doi.org/10.5281/zenodo.24648313}{10.5281/zenodo.24648313}}.

\section{Related Work}
\label{sec:related}

\textbf{Network emulation platforms.} Mininet~\cite{mininet}, Containernet~\cite{containernet} and MaxiNet~\cite{maxinet} established network namespaces as a practical substrate for large emulated topologies. LEO-specific platforms add orbital dynamics: StarryNet~\cite{starrynet} and OpenSN~\cite{opensn} emulate constellations with containers running real routing software, and Hypatia~\cite{hypatia} and StarPerf~\cite{starperf} study LEO networks through simulation. SplitNN~\cite{splitnn} and NetLoom~\cite{netloom} accelerate the construction of virtual networks with thousands of nodes on one host. These systems mainly optimize topology setup and orchestration; distributed emulators such as CORE~\cite{core} scale out across hosts. The runtime path that installs already computed routes has received less attention, and IsoRoute targets that path.

LFRI~\cite{lfri} is a recent study of this path. It identifies route installation as a scaling limit in container-based emulation of LEO mega-constellations and, like IsoRoute, moves forwarding state into eBPF maps. Its design filters ``ineffective installations'', so that after a route recomputation only routes that differ from the kernel table are installed, and it connects to the routing suite through an SDK or a Netlink adaptation layer. An eBPF forwarder handles link- and network-layer processing, and the evaluation reports installation latency and path availability. LFRI attributes the cost mainly to contention on a global lock and the resulting cache invalidation.

Our measurements in \S\ref{sec:rootcause} lead to a different emphasis. Once writers are serialized under \texttt{rtnl\_mutex}, adding writers barely changes throughput, whereas idle Netlink subscribers in unrelated namespaces lower the ceiling: every route update scans the multicast subscriber list inside the critical section. IsoRoute therefore keeps FRR's incremental installation unchanged and reduces the cost of each installation rather than the number of installations. It takes routes from the FPM export interface with a one-line \texttt{zebra} change, while LFRI intercepts them at Netlink. The two designs also place the dataplane boundary differently. IsoRoute bypasses only the FIB lookup and reuses the kernel's existing forwarding functions: TTL decrement, ICMP errors, neighbor resolution through \texttt{bpf\_redirect\_neigh()}, \texttt{netem} link delays on the egress qdisc, and locally originated traffic routed to \texttt{dummy0}. Per-node ordering and an atomic table swap on \texttt{zebra} reconnection keep the eBPF tables consistent, and we check fidelity by comparing converged forwarding tables entry by entry and by preserving configured link delays (\S\ref{sec:dataplane}). The implementation, topology scripts and measurement scripts are available as described in \S\ref{sec:availability}.

\textbf{Kernel networking scalability.} Contention on the RTNL mutex has been discussed in the Linux networking community for years, including talks and patch series that move rtnetlink toward per-namespace locking and per-namespace device unregistration~\cite{pernetns-rtnl,pernetns-rtnl-patch,pernetns-unregister}. These changes reduce serialization across namespaces; scanning the multicast subscriber list in the notification path is a separate cost. Earlier analyses of Linux multicore scalability~\cite{clements-scalable,boyd-wickizer-analysis} examine lock contention and cache effects in general; we characterize how serialization and subscriber scanning interact under namespace-heavy route churn.

\textbf{Routing offload and programmable dataplanes.} FRR's FPM interface~\cite{frr-fpm}, used by SONiC~\cite{sonic} to program switch ASICs, and mechanisms such as gRIBI~\cite{gribi} export routes to forwarding planes outside the kernel. IsoRoute uses FPM for a different purpose: removing a shared kernel bottleneck in emulation. eBPF systems including XDP~\cite{xdp}, Cilium~\cite{cilium}, Katran~\cite{katran} and Calico~\cite{calico-ebpf} show that programmable kernel datapaths can forward efficiently; they target policy enforcement, load balancing and connection tracking rather than per-node forwarding state for thousands of routing instances.

\section{Conclusion}
\label{sec:conclusion}

Container-based LEO network emulation hits a runtime scaling limit: every route update scans, inside \texttt{rtnl\_mutex}, a Netlink subscriber list shared by all namespaces, so kernel installation capacity falls while the offered load grows superlinearly with constellation size; on our testbed the two cross at about 1,200--1,300 containers. IsoRoute exports routes over FPM, writes forwarding state into one eBPF table per node, and bypasses only the kernel's FIB lookup while leaving everything else to the kernel, removing this coupling without changing the control plane. From 500 to 3,000 FRR containers it sustains about $3\times10^5$ installations per second with no decline with scale, and it cuts the OSPF data-plane convergence time of a 1,584-node Starlink Shell-1 topology from 594.7\,s to 49.5\,s. Whether other multi-tenant settings that rely on rtnetlink, such as large-scale CNI link setup, show the same coupling is a question for future work.

\bibliographystyle{plain}
\bibliography{references}

\appendix
\section{Configuration and Forwarding Program}
\label{sec:appendix}

\subsection{Zebra Dataplane Settings}

The baseline kernel path uses FRR 9.1 with the default \texttt{zebra} dataplane configuration. The \texttt{zebra} daemon runs in each container namespace. Only the parameters that were actually set are listed below:

\begin{itemize}[leftmargin=1.5em]
\item Dataplane queue depth: FRR 9.1 default, 200 pending updates (\texttt{zebra dplane limit})~\cite{frr-src}
\item Route installation threads: 1 (single-threaded netlink writer)
\item Netlink socket buffer: system default
\item Route processing mode: asynchronous (\texttt{zebra} queues routes internally before netlink submission)
\end{itemize}

\subsection{System Configuration}

All experiments use the following system parameters:

\begin{itemize}[leftmargin=1.5em]
\item File descriptor limit: \texttt{ulimit -n 1048576}
\item Netlink socket receive buffer: \texttt{net.core.rmem\_default = 4194304}
\item Netlink socket maximum buffer: \texttt{net.core.rmem\_max = 16777216}
\item Container orchestration: Docker with one network namespace per node
\item IRQ balancing: enabled (default kernel configuration)
\end{itemize}

No kernel patches or non-standard networking stack modifications were applied. The baseline uses unmodified FRR; IsoRoute uses the same FRR build plus the one-line \texttt{zebra} patch that disables kernel route installation.

\subsection{eBPF Forwarding Program}
\label{sec:ebpf-code}

The listing below gives the simplified \texttt{tc} program that implements the data plane of \S\ref{sec:design-forward}. The program is attached at both the ingress of inter-satellite links and the egress of \texttt{dummy0}; it selects the node's table by the netns cookie, looks up the destination prefix, and redirects a hit to the next hop. Packets it does not handle return \texttt{TC\_ACT\_OK} and continue through the kernel, so kernel forwarding remains responsible for local delivery, ICMP errors, neighbor resolution, and link delays.

\begin{lstlisting}[style=routercode]
SEC("classifier")
int router_tc(struct __sk_buff *skb)
{
    __u32 saddr = 0, daddr = 0;
    struct iphdr *iph = parse_ipv4(skb, &saddr, &daddr);
    if (!iph)                    /* IPv6, ARP, non-IP */
        return TC_ACT_OK;
    if (is_multicast(daddr) || iph->protocol == 89)
        return TC_ACT_OK;        /* control traffic */

    struct cookie_key ck = {
        .cookie = get_dev_netns_cookie(skb) };
    rid_t *ridp = bpf_map_lookup_elem(&cookie2rid, &ck);
    if (!ridp)
        return TC_ACT_OK;
    void *fib_tbl = bpf_map_lookup_elem(&fib_dir, ridp);
    if (!fib_tbl)
        return TC_ACT_SHOT;

    struct router_fib_key key = {
        .prefixlen = 32, .addr = daddr };
    struct router_fib_info *fib =
        bpf_map_lookup_elem(fib_tbl, &key);
    if (!fib)                    /* route miss */
        return TC_ACT_OK;        /* -> kernel */
    if (skb->ingress_ifindex) {  /* transit, not local */
        if (iph->ttl <= 1)       /* TTL expiry */
            return TC_ACT_OK;    /* -> kernel sends ICMP */
        dec_ttl(iph);            /* incremental checksum */
    }
    struct bpf_redir_neigh nh = {
        .nh_family = AF_INET, .ipv4_nh = fib->gw_ip };
    return bpf_redirect_neigh(fib->ifindex, &nh,
                              sizeof(nh), 0);
}
\end{lstlisting}

\end{document}